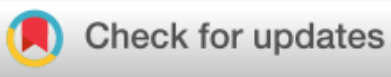

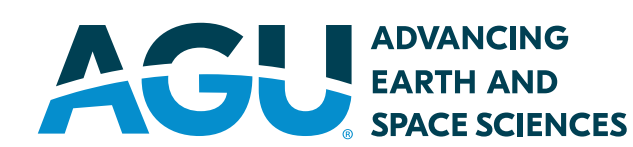

**Key Points:**

- We present new three-dimensional model calculations that predict a longer lifetime for Earth's biosphere than previous one-dimensional models
- We suggest that some land and aquatic plants could survive at lower carbon dioxide levels than the canonical 10 parts per million limit
- Earth's photosynthetic biosphere could survive for another 1.8 billion years, approaching the limit at which Earth would lose its oceans

**Correspondence to:**
J. Haqq-Misra,
jacob@bmsis.org

**Citation:**
Haqq-Misra, J., & Wolf, E. (2026). Maximum lifetime of the vegetative biosphere. *Journal of Geophysical Research: Atmospheres*, *131*, e2025JD045586. https://doi.org/10.1029/2025JD045586

Received 6 OCT 2025
Accepted 21 MAY 2026

**Author Contributions:**
**Conceptualization:** Jacob Haqq-Misra
**Data curation:** Jacob Haqq-Misra
**Methodology:** Jacob Haqq-Misra
**Software:** Eric Wolf
**Writing – original draft:** Jacob Haqq-Misra
**Writing – review & editing:** Eric Wolf

# Maximum Lifetime of the Vegetative Biosphere

**Jacob Haqq-Misra[1] and Eric Wolf[1,2]**

[1]Blue Marble Space, Seattle, WA, USA, [2]Laboratory for Atmospheric and Space Physics, University of Colorado Boulder, Boulder, CO, USA

**Abstract** We use a three-dimensional model to calculate steady-state climates at various intervals in Earth's future, across a parameter space of increasing insolation and decreasing $CO_2$ mixing ratio. Comparison with prior results shows an overestimation of warming by one-dimensional models when solar constant is increased and $CO_2$ mixing ratio is fixed. We consider two future trajectories as limiting cases: strong weathering, in which surface temperature remains constant but $CO_2$ is drawn down; and weak weathering, in which $CO_2$ remains constant and surface temperature increases. Under strong weathering, we find the conventional 10 ppm $CO_2$ starvation limit for C4 photosynthesis that occurs at 1.35 Gyr; however, we suggest that crassulacean acid metabolism (CAM) photosynthesis could persist below this limit and note that aquatic macrophytes can utilize dissolved bicarbonate if atmospheric $CO_2$ is low. If we take the $CO_2$ starvation limit at 1 ppm instead, then the vegetative biosphere could continue until 1.84 Gyr. Thermal limits apply instead under weak weathering, in which Earth would be too hot for most land plants at 1.68 Gyr (>323 K) and too hot for all land plants (>338 K) at 1.87 Gyr. These lifetimes approach the moist and runaway greenhouse limits for Earth. We discuss other possible mechanisms for extending the lifetime of Earth's biosphere, noting that both technological intervention and evolutionary processes could enable life to adapt to a brightening sun.

**Plain Language Summary** The ultimate life span of Earth's biosphere is limited due to the steady brightening of the sun as it progresses in age. Earth's long-term carbon cycle may respond by drawing carbon dioxide out of the atmosphere and into carbonate rocks, thereby reducing the greenhouse effect and offsetting the increased sunlight. Most prior studies have argued that this would eventually make carbon dioxide levels too low to sustain photosynthesis, thus marking the end of the biosphere as we know it. In this study, we use a three-dimensional computational climate model to calculate scenarios of Earth's future climate with increasing sunlight and decreasing carbon dioxide. We show that Earth's biosphere could survive for much longer than indicated in most studies, noting that some photosynthetic life on Earth can thrive at very low carbon dioxide levels. We also explore possible trajectories in which carbon dioxide does not decrease but instead remains constant. Earth's vegetative biosphere could survive up to about 1.8 billion years from now, about the same time that Earth would lose its oceans to space. We discuss other ways that biological evolution as well as technological intervention could even further extend the lifetime of Earth's vegetative biosphere.

## 1. Introduction

How long will life on Earth survive? This question was once the exclusive domain of theologians and philosophers, but today it represents a tractable area of study among astrobiologists. The ultimate fate of Earth is predicted from models of stellar evolution, in which the sun expands into its red giant phase ∼5 Gyr from now, swelling up past the orbit of Earth and engulfing our planet. If the sun were to lose sufficient mass during its entry into the red giant phase, then Earth might alternatively be ejected from its current orbit and displaced to wander as an unbound "rogue" planet. But most projections of Earth's ultimate fate tend to favor an outcome where, in the words of Robert Frost, "the world will end in fire." Models of stellar evolution (e.g., Goldstein, 1987; Schröder & Connon Smith, 2008) predict that main sequence stars, like our sun, brighten steadily with time: the process of hydrogen fusion causes stars to contract as they build up denser cores, which increases the rate of fusion reactions (e.g., Baraffe et al., 2015). For the early stages of the solar system, the sun was 20%–30% less luminous, which has led to intense study of the "faint young sun" paradox for early Earth and early Mars (e.g., Donn et al., 1965; Feulner, 2012; Goldblatt & Zahnle, 2011; Kasting, 2010; Sagan & Mullen, 1972). For Earth's future, these models of stellar evolution predict that Earth will receive a steady increase in incident radiation as the sun ages. Even before Earth is engulfed by the sun's expansion, this steady brightening could cause sufficient climate warming that life on Earth becomes impossible.

The discovery of stabilizing feedback loops in the Earth system gave a mechanism for understanding the relationship between plate tectonics and planetary habitability. Specifically, the regulation of Earth's long-term climate by the carbonate-silicate cycle provided a way of predicting the evolution of Earth's climate as the sun ages (Walker et al., 1981). The carbonate-silicate cycle describes the process by which carbon dioxide ($CO_2$) is cycled over geologic time. The cycle begins as $CO_2$ is outgassed from volcanoes and enters the atmosphere, where it can serve as a greenhouse gas. As $CO_2$ is dissolved in rainwater and reaches the surface, it reacts with silicate rocks to form calcium and bicarbonate ($HCO_3$) ions that runoff into the ocean. Marine calcifying organisms use these ions to construct shells and exoskeletons, which precipitate to the seafloor as calcium carbonate and are subducted into Earth's mantle. (In the absence of biology, calcium carbonate would still precipitate to the seafloor after exceeding the saturation threshold (Walker et al., 1991).) The high temperature and pressure of the mantle causes the carbonate rocks to metamorphically convert back into silicate rocks, which releases $CO_2$ through volcanic outgassing—thus completing the cycle. The rate of weathering of silicate rocks by rainwater depends on temperature: silicate weathering would increase in a warmer climate, which would draw down more $CO_2$ from the atmosphere. This carbonate-silicate cycle operates on a $\sim$0.5 Myr timescale as a regulator of Earth's atmospheric carbon dioxide through deep time (Berner et al., 1983).

Projections of the carbonate-silicate cycle into Earth's future raised further speculation as to whether or not the biosphere as we know it could survive as long as the planet. The study by Lovelock and Whitfield (1982) used a simple zero-dimensional (0-D) climate model to find that Earth's photosynthetic biosphere would end 100 Myr from now. The authors reasoned that the carbonate-silicate cycle would draw atmospheric $CO_2$ levels down to 150 parts per million (ppm) by this time, which represents the critical level for C3 photosynthesis. C3 photosynthesis is the most common carbon fixation metabolic pathway and represents a large fraction of Earth's vegetation (95%). Lovelock and Whitfield (1982) took this C3 photosynthesis threshold to mark the end of the biosphere as a whole. Caldeira and Kasting (1992) revisited this problem and argued that Earth's biosphere should persist for another 0.9–1.5 Gyr from now. Using a more sophisticated 0-D climate model, Caldeira and Kasting (1992) demonstrated that there are two important limits to consider to the future of life on Earth. The first is the increase in global temperature that should occur as the sun steadily brightens, with 1.5 Gyr calculated as the time at which global temperature exceeds 323 K (and an even harder limit calculated to occur $\sim$1 Gyr after this with the loss of Earth's oceans to space.) However, Caldeira and Kasting (1992) also noted that C4 photosynthesis could thrive at $CO_2$ levels as low as 10 ppm. C4 photosynthesis is an adaptation of the C3 carbon fixation pathway where the plant improves its photosynthetic efficiency by reducing photorespiration. C4 plants represent 3% of Earth's vegetation and include crops such as corn, sugar cane, and cabbage as well as daisies, grasses, and weeds. The updated estimate of the lifetime of the biosphere by Caldeira and Kasting (1992) suggested that Earth could sustain a C4-based biosphere for 0.9 Gyr from now.

A number of expansions to this story have appeared in the literature, but this basic narrative developed by Caldeira and Kasting (1992) remains the prevailing paradigm among astrobiologists. Subsequent investigations conducted by Franck et al. (2000) and Lenton and von Bloh (2001) applied more sophisticated parameterizations into the 0-D climate model of Caldeira and Kasting (1992). The advanced parameterization of biosphere productivity in the Franck et al. (2000) study gave a shorter biosphere lifetime of about 500 Myr. The study by Lenton and von Bloh (2001) incorporated stronger biotic weathering effects, which resulted in a longer lifetime of 800 Myr. Later studies conducted by Rushby et al. (2018) and de Sousa Mello and Friaça (2020) also applied more complex one-dimensional (1-D) climate models to calculate the lifetime of the biosphere. Rushby et al. (2018) developed a more sophisticated biogeochemical model, coupled to a 1-D radiative transfer equilibrium climate model, which enabled an exploration of the dependence of this end of the biosphere lifetime on planetary mass. The logical extension of the C4 photosynthesis limit, as Rushby et al. (2018) emphasized, is the end of a large-scale photosynthetic biosphere at 10 ppm $CO_2$ with an extended phase of microbial habitability until the planet becomes too hot to host even extremophiles. Another study by de Sousa Mello and Friaça (2020) constructed a comparison of estimates found in the literature and, drawing upon their own 1-D climate model results, estimated the end of C4 photosynthesis at $840 + 270/-100$ Myr. This range describes the underlying uncertainty regarding weathering rates and other disagreements among model parameterizations, but even the upper limit of this range falls short of the $\sim$1.5 Gyr (according to Caldeira and Kasting (1992)) or more until Earth's surface begins to lose water and becomes truly uninhabitable.

Finally, recent work by Graham et al. (2024) developed improved models for the productivity of C3 and C4 plants as well as silicate weathering, which were coupled to the Caldeira and Kasting (1992) climate model to demonstrate an expanded lifetime of the photosynthetic biosphere. Using recent data, Graham et al. (2024) argued that C3 photosynthesis can remain viable at $CO_2$ mixing ratios as low as 2.9 ppm, rather than the conventional value of 10 ppm assumed by Caldeira and Kasting (1992) and others. This revised limit for C3 planets by itself would extend the Caldeira and Kasting (1992) estimate of the lifetime of the biosphere from 0.9 to 1.3 Gyr. However, Graham et al. (2024) also suggested that recent data indicate a relatively weak dependence of silicate weathering on temperature, which would suggest that $CO_2$ may not necessarily continuously decrease in the future due to a brightening sun. Indeed, the model developed by Graham et al. (2024) demonstrated a more complex trajectory for the future evolution of atmospheric $CO_2$, with only a modest net decrease over a ~1.5 Gyr time span when compared with the assumption of a strong weathering dependence on temperature made by Caldeira and Kasting (1992) and others. If silicate weathering is as weakly temperature dependent as Graham et al. (2024) suggested, then the lifetime of Earth's vegetative biosphere will be limited by overheating rather than $CO_2$ starvation. Graham et al. (2024) placed this limit at 1.6–1.86 Gyr from now, depending on whether the threshold for plant productivity is taken to be the traditional limit of 323 K or the higher limit of 338 K. Graham et al. (2024) noted that this extended lifetime of Earth's biosphere also falls within the range of predictions for Earth to lose its oceans through a moist or runaway greenhouse state (predicted to be at 1.5 Gyr by the 0-D model of Caldeira and Kasting (1992), 1.1 Gyr by the 3-D model of Leconte et al. (2013), and 2.19 Gyr by the 3-D model of Wolf and Toon (2015)). If weathering remains strongly temperature dependent, then $CO_2$ starvation could still limit the biosphere at an earlier time; but if weathering is weakly temperature dependent, then the biosphere may remain viable as long as Earth's oceans.

This study examines the problem of Earth's biosphere by considering strong and weak temperature dependence of silicate weathering as two limits, using a three-dimensional (3-D) general circulation model. Such models are much more computationally expensive than simpler 1-D models, but 3-D models provide numerous advantages that include calculation of temperature, wind, and other atmospheric properties across latitude, longitude, and altitude; explicit representation of the hydrological cycle and ice formation; spatially resolved spectral radiative transfer calculations; and parameterizations of clouds and other physical processes. Calculations with the Exo-CAM 3-D climate model are first compared with the lineage of 0-D and 1-D models used by Caldeira and Kasting (1992) and others (see Table 1 by Graham et al. (2024) for a summary), which reveal limitations in the ability of simplified models to represent climate responses to large increases in insolation as well as significant decreases in $CO_2$ mixing ratio (§3). We next analyze a set of 29 ExoCAM calculations across a parameter space of varying instellation and $CO_2$ mixing ratio, using a habitability metric to demonstrate changes in the biosphere (§4). We then calculate the maximum lifetime of the biosphere under the limiting cases of strong and weak weathering (§5) and we discuss these limits in the context of the moist greenhouse, runaway greenhouse, and other possibilities for Earth's future (§6).

## 2. Model Descriptions

The sections that follow utilize several different models. Some readers may be familiar with some or all of these models, while others may not have any prior knowledge of these models. The *dimensionality* of the various models compared is the primary focus, so readers unfamiliar with these models may nevertheless proceed with the sections that follow. We have chosen to provide descriptions of the models used in the following calculations in the Appendix, allowing readers to selectively refer to these sections as needed. (Readers who prefer to read a full technical methods section are recommended to see the Appendix at this time.)

We perform calculations using the Caldeira and Kasting (1992) 0-D energy balance model (Appendix A) as well as the Williams and Kasting (1997) and Haqq-Misra and Hayworth (2022) 1-D energy balance models (Appendix B), all of which use pre-computed parameterizations for the radiative forcing due to $CO_2$. We also present several sets of calculations from the ExoCAM 3-D general circulation model (Appendix C), which include previously published results by Wolf and Toon (2015) and Wolf et al. (2018) as well as a set of new calculations for this present study. Calculations with ExoCAM explicitly resolve radiative forcing due to $CO_2$ across latitude, longitude, and altitude. Our configuration of ExoCAM in this study is intended to facilitate as direct comparison as possible with these previously published results. We use a habitability metric developed by Woodward et al. (2025) for examining spatial habitability changes in these ExoCAM simulations (Appendix D).

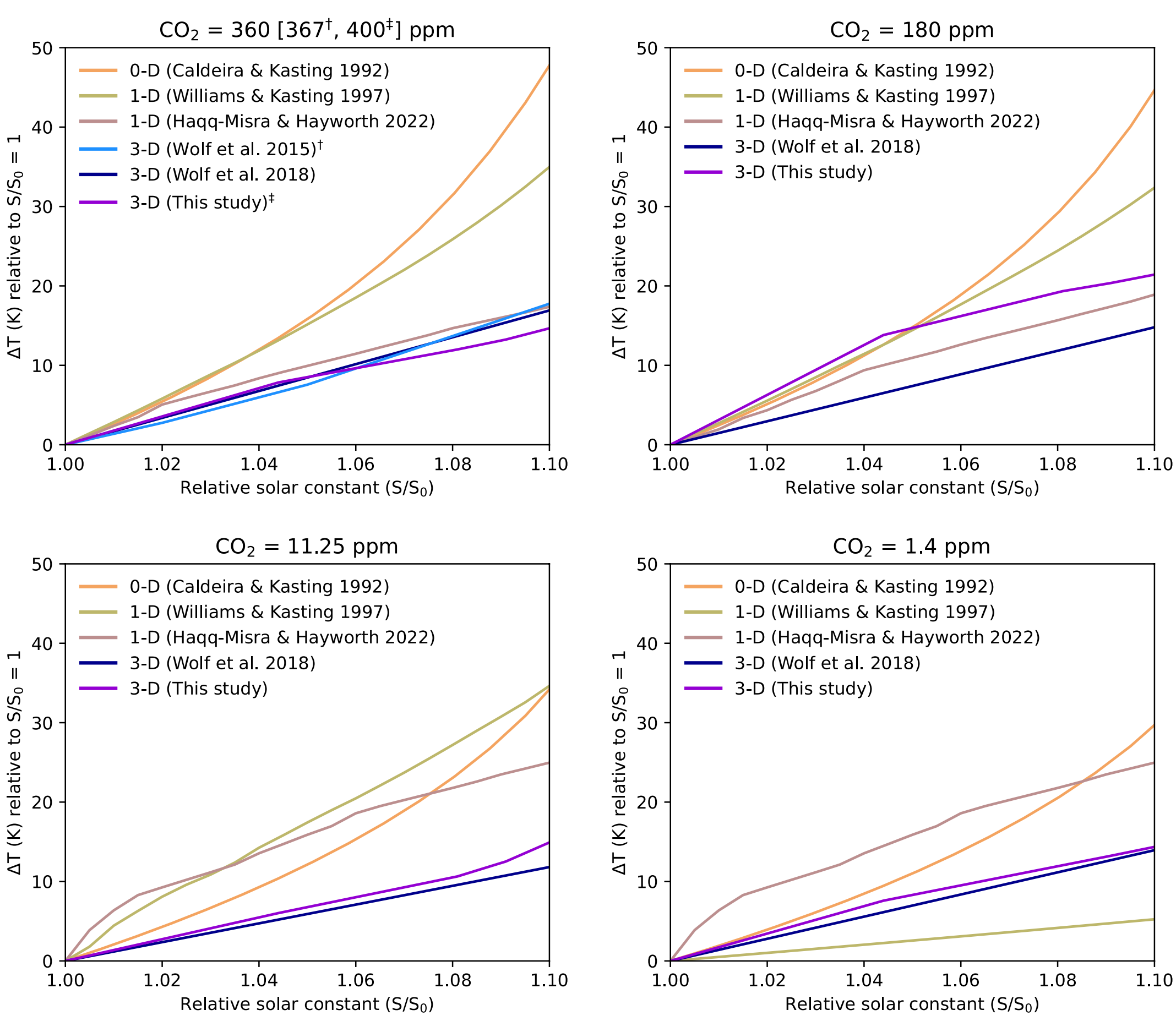

**Figure 1.** Simplified models give different results from three-dimensional models for the expected temperature change due to an increased solar constant. Calculations assume a fixed $CO_2$ mixing ratio, with panels showing four different values ranging from present-day Earth (400 ppm for this study, 360 ppm or 367 ppm in other studies) to future Earth scenarios in which $CO_2$ decreases due to weathering. Simplified energy balance climate models with zero dimensions (orange) or one dimension in latitude (olive, brown) tend to show larger temperature changes than those calculated with more complex three-dimensional general circulation models (blue, indigo, violet).

## 3. Model Intercomparison

The future climate of Earth will evolve in response to a brightening sun and the carbonate-silicate cycle. Exploring this behavior with a climate model thus requires simultaneously adjusting the incident stellar radiative flux and the $CO_2$ mixing ratio. This is the approach that was followed by Caldeira and Kasting (1992) and many subsequent studies. It is worth emphasizing that many of these subsequent studies have either utilized the simplified zero-dimensional (0-D) model of Caldeira and Kasting (1992) in tandem with more sophisticated models of silicate weathering and biological productivity (e.g., Franck et al., 2000; Lenton & von Bloh, 2001; Graham et al., 2024), or have relied on the one-dimensional (1-D) climate model parameterizations of Williams and Kasting (1997) instead (e.g., de Sousa Mello & Friaça, 2020; Ozaki & Reinhard, 2021). The models of Caldeira and Kasting (1992) and Williams and Kasting (1997) likewise derive from a common lineage (e.g., Kasting et al. (1993), with the more recent version of Kopparapu et al. (2013) utilized by Rushby et al. (2018)), so these calculations cannot be considered fully independent from one another. This underscores the importance of model intercomparison studies that examine the extent to which such simplified climate models compare with more complex three-dimensional (3-D) climate models, in this case for situations involving increasing insolation and decreasing $CO_2$. Because this problem spans a two-parameter space, we investigate changes in each parameter separately, first varying insolation while holding $CO_2$ constant (Figure 1), and then varying $CO_2$ mixing ratio while holding insolation constant (Figure 2). All calculations otherwise assume present-day Earth conditions with a 1 bar $N_2$ background atmosphere and variable water vapor content.

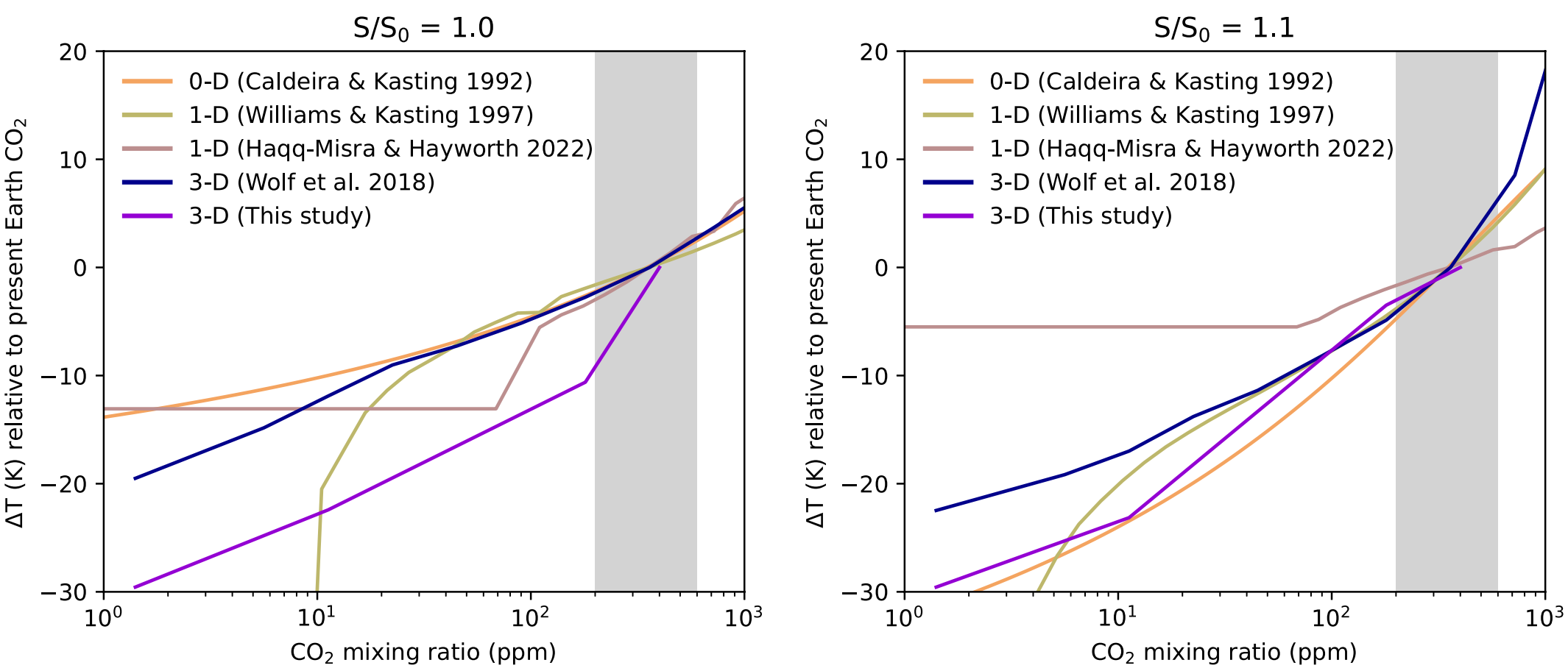

**Figure 2.** Simplified models give different results from three-dimensional models for the expected temperature change due to an increased $CO_2$ mixing ratio. Calculations assume a fixed relative solar constant, with panels showing values for present-day solar flux (left, $S/S_0 = 1$) and future solar flux at ∼1.0 Gyr (left, $S/S_0 = 1.1$). Simplified energy balance climate models with zero dimensions (orange) or one dimension in latitude (olive, brown) show greater agreement with more complex three-dimensional general circulation models (indigo, violet) at $CO_2$ values with small deviations from present-day Earth (i.e., ∼200–600 ppm, gray shading) and larger deviations at lower $CO_2$ values. Some cases reveal model limitations in the ability for some one-dimensional models (olive, brown) to capture responses to changes in $CO_2$ lower than 10 ppm.

The calculations shown in Figure 1 compare the global average temperature change given by several different models as solar constant ($S/S_0$) increases, relative to each model's value of global average temperature at $S/S_0 = 1$. All calculations are performed at fixed values of $CO_2$ mixing ratio, which include present-day Earth values (upper left; 400 ppm for this study, 360 ppm or 367 ppm in other studies), 180 ppm (upper right), 11.25 ppm (lower left), and 1.4 ppm (lower right). The different present-day Earth values reflect that each model required a different configuration in order to obtain a present-day Earth steady-state with a global average temperature of 288 K. We ran additional cases at $CO_2$ mixing ratios of 360 and 367 ppm, but these were 1–2 K cooler than 288 K; this is consistent with other recent simulations using this updated version of ExoCAM that used 500 ppm for its present-day Earth case (Deitrick et al., 2025). The values of 180, 11.25, and 1.4 ppm are selected in particular to facilitate direct comparisons with the values that were used in the published results from the 3-D ExoCAM model by Wolf and Toon (2015) and Wolf et al. (2018). The value of 180 ppm is near the threshold for C3 photosynthesis, the 11.25 ppm value is about at the threshold for C4 photosynthesis, and the 1.4 ppm corresponds to an even lower limit for photosynthesis (discussed in §5). We performed additional simulations at 10 and 1 ppm that showed insignificant differences with the comparable cases at 11.25 ppm and 1.4 ppm. We note that our choice of these specific values has no effect on our later calculations of the lifetime of Earth's biosphere.

The 0-D model of Caldeira and Kasting (1992) (orange curve) gives consistently larger temperature changes in response to increases in solar constant compared to all other models. The 1-D model of Williams and Kasting (1997) (olive curve) likewise gives larger temperature changes than others, comparable to but slightly lower than Caldeira and Kasting (1992) across much of the calculation space; the exception is at very low (1.4 ppm) $CO_2$, where the Williams and Kasting (1997) model gives lower temperature changes than other models. A more recent version of a 1-D model in the lineage of Williams and Kasting (1997) was developed by Haqq-Misra and Hayworth (2022) (brown curve), which shows more modest changes in temperature at $CO_2$ values of 360 and 180 ppm; however, the Haqq-Misra and Hayworth (2022) model is more comparable to the Caldeira and Kasting (1992) model at lower levels of $CO_2$ (11.25 and 1.4 ppm).

Additional results in Figure 1 show calculations with the ExoCAM 3-D climate model (see Appendix C and Wolf et al. (2022) for a complete model description). Calculations from the study by Wolf and Toon (2015) (blue curve) are limited to the case of present-day $CO_2$ values (367 ppm), while the results of the study by Wolf et al. (2018) (indigo curve) are included in all four panels showing different $CO_2$ mixing ratios. These previously published 3-D results generally agree with each other for the present-day Earth case (upper left) but differ significantly with the 0-D and 1-D models. The 3-D calculations compare reasonably with the Haqq-Misra and Hayworth (2022) model at present-day Earth and 180 ppm $CO_2$ mixing ratios, but otherwise the 3-D models generally show a more modest relationship between $S/S_0$ and $CO_2$ mixing ratio than the simpler 0-D and 1-D models. Such a result is not necessarily surprising, as 3-D models are capable of representing temporally and spatially varying properties of climate such as cloud cover, fractional ocean coverage, active hydrological cycles, and other features that must be parameterized or eliminated in simpler models. Another contributing factor is that the ExoRT radiative transfer package used by ExoCAM represents a separate model linage with independent development from the Kasting et al. (1993) lineage, so some aspects of these differences could result from the use of different methodologies and assumptions in the representation of radiative transfer. An initial implication of these results is that any prior studies that utilized the simpler models of Caldeira and Kasting (1992), Williams and Kasting (1997), or others in this lineage may have overestimated the warming that would occur in response to a brightening sun.

**Table 1**
*Summary of the 29 ExoCAM Simulations Conducted in This Study Across a Parameter Space of Relative Solar Constant ($S/S_0$) and $CO_2$ Mixing Ratio, With an Otherwise Earth-like Atmosphere With 1 bar $N_2$ and 1.8 ppm $CH_4$, Earth-Like Topography, and a Slab Ocean*

| Scenario | $S/S_0$ | $CO_2$ (ppm) | Surface temperature (K) |
|---|---|---|---|
| Present-day | 1.0 | 400 | 287.9 |
| | 1.0 | 180 | 277.3 |
| | 1.0 | 11.25 | 265.5 |
| | 1.0 | 1.4 | 258.4 |
| 0.5 Gyr Future | 1.044 | 400 | 295.8 |
| | 1.044 | 180 | 291.1 |
| | 1.044 | 135 | 288.6 |
| | 1.044 | 11.25 | 271.5 |
| | 1.044 | 1.4 | 265.9 |
| 0.9 Gyr Future | 1.081 | 400 | 299.9 |
| | 1.081 | 180 | 296.7 |
| | 1.081 | 11.25 | 276.2 |
| | 1.081 | 1.4 | 270.4 |
| 1.0 Gyr Future | 1.091 | 400 | 301.2 |
| | 1.091 | 180 | 297.7 |
| | 1.091 | 34 | 288.5 |
| | 1.091 | 11.25 | 278.0 |
| | 1.091 | 1.4 | 271.6 |
| 1.5 Gyr Future | 1.143 | 400 | 309.4 |
| | 1.143 | 180 | 303.8 |
| | 1.143 | 11.25 | 291.9 |
| | 1.143 | 6.0 | 288.6 |
| | 1.143 | 1.4 | 278.1 |
| 2.0 Gyr Future | 1.2 | 400 | 348.3 |
| | 1.2 | 180 | 344.9 |
| | 1.2 | 11.25 | 300.2 |
| | 1.2 | 1.4 | 292.9 |
| | 1.2 | 0.45 | 288.0 |
| | 1.2 | 0 | 280.1 |

Finally, new calculations performed for this study (listed in Table 1) with the most recent version of ExoCAM (purple curve) show general agreement with previous ExoCAM results, with the exception of a higher predicted temperature change for the case with a 180 ppm $CO_2$ mixing ratio. One reason for these differences is that the ExoRT radiative transfer package was upgraded in 2020, which included migrating the source of molecular absorption data from the HITRAN 2004 spectroscopic database to HITRAN 2016. Some of the differences between the new ExoCAM calculations and those previously published by Wolf and Toon (2015) and Wolf et al. (2018) are found (see discussion below). The full set of 0-D, 1-D, and 3-D models in Figure 1 gives an appropriate demonstration of a maxim articulated by Box (1976): "Since all models are wrong the scientist must be alert to what is importantly wrong. It is inappropriate to be concerned about mice when there are tigers abroad." In this case, the mousy variations among the 3-D models themselves are far less alarming than the tigerish differences that arise from comparing 0-D/1-D models with 3-D models.

The calculations shown in Figure 2 compare the global average temperature change given by several different models as $CO_2$ mixing ratio increases, relative to each model's value of global average temperature at present-day Earth $CO_2$ levels. All calculations are performed at fixed values of solar constant, with $S/S_0 = 1.0$ (left) corresponding to present-day solar flux and $S/S_0 = 1.1$ to future solar flux at ∼1 Gyr. In these cases the 0-D model of Caldeira and Kasting (1992) (orange curve) compares reasonably well with the 3-D models of Wolf et al. (2018) for $S/S_0 = 1.0$ and the model calculations of this study at $S/S_0 = 1.1$. The 1-D model of Williams and Kasting (1997) (olive curve) behaves consistently with the 0-D and 3-D models down to a $CO_2$ mixing ratio of ∼10 ppm, below which the model shows very large decreases in temperature that indicate global glaciation. The 1-D model of Haqq-Misra and Hayworth (2022) similarly compares well with the 0-D and 3-D models down to a $CO_2$ mixing ratio of ∼100 ppm, below which the model shows no additional cooling due to decreases in $CO_2$. (This is most likely an artifact of the sparse grid of low-$CO_2$ points in the radiative transfer lookup table developed by Haqq-Misra and Hayworth (2022).) The 3-D results of Wolf et al. (2018) show the same slope as the model calculations of this study, but with an offset of ∼10 K for $S/S_0 = 1.0$ and slightly less for $S/S_0 = 1.1$ at lower $CO_2$ mixing ratios. These differences become less for higher $CO_2$ amounts and warm global mean temperatures. The set of results shown in Figure 2 indicate that not all models perform equally well at low $CO_2$ mixing ratios (which is perhaps understandable, given that much of contemporary climate modeling focuses on understanding the response to *higher* levels of $CO_2$). For this set of experiments in isolation, the Caldeira and Kasting (1992) appears comparable to the 3-D model results, all of which are preferable to the shown 1-D results.

Both the older (e.g., Wolf and Toon (2015); Wolf et al. (2018)) and updated (e.g., Wolf et al. (2022) and this work) forms of ExoCAM share core surface, water cycle, and atmospheric dynamics routines. A key technical difference in the code base lies in the radiative transfer which was completely overhauled by increasing the spectral resolution, updating absorption data (HITRAN 2004 to HITRAN 2016), using a new gas overlap scheme, and employing a modern treatment of $CO_2$ absorption (Perrin & Hartmann, 1989; Wordsworth et al., 2010). Furthermore, the updated version of ExoCAM assumes a time-invariant parameterization of heat transport within the slab ocean, versus including a seasonally varying parameterization in earlier models. ExoCAM now also assumes a 10% lower snow albedo compared to the earlier versions. We point out these potential triggers for differences, but here we refrain from the tedious dive into teasing out changes beget by a decade of code base evolution. Note also the recent work performed by Deitrick et al. (2025), who performed benchmark calculations of ExoCAM with oxygen species for modern Earth conditions.

As a final remark, we performed an additional simulation at 2.0 Gyr ($S/S_0 = 1.2$) with the $CO_2$ mixing ratio set to zero (Table 1, bottom row). This simulation remains above-freezing with a 280 K average surface temperature, in part due to the greenhouse effect of methane (1.8 ppm), which is ~8 K cooler than the comparable simulation with a 0.45 ppm $CO_2$ mixing ratio. The magnitude of warming obtained from such a small amount of $CO_2$ is not necessarily surprising, given that these simulations do not include any other molecular species that overlap with the $CO_2$ absorption bands. Nevertheless, it is instructive to note such sensitivities and perhaps perform future comparisons with other 3-D models at low and zero values of $CO_2$ mixing ratio, given that radiative transfer models at low $CO_2$ and high insolation may give a greater range of predictions compared to present-day conditions.

This exercise once again demonstrates the importance of comparing models of different complexities and lineages to one another. In some cases, we find general trends that persist across multiple models, while in other cases, we find significant divergence in results. When models do not agree, this can indicate many possibilities. One position is that more complex models tend to be more reliable because they are able to represent a wider range of physical processes with greater completeness and interconnection. To an extent, the 3-D model results shown in Figures 1 and 2 differ from the 0-D and 1-D models due to the ability of the 3-D models to represent non-linear processes with temporal and spatial variation, which generally compares better with the known properties of Earth's climate than simpler models. This position would suggest that the 3-D models are more indicative of how Earth's climate would actually behave. But another position is that disagreements among models tend to reveal deep uncertainties about the most meaningful processes driving climate for the problem of interest. Such a position could argue that simple models can often provide high confidence results under certain constraints while also being easier to understand. This returns us again to the wisdom of Box (1976): “Since all models are wrong the scientist cannot obtain a ‘correct’ one by excessive elaboration.” We do not purport to claim that the results of ExoCAM (before or after the updates in 2020) represent “truth” in any sense of the word, nor do we claim that 3-D models are always inherently more accurate or otherwise “better” than simpler models. But we nevertheless note that the trends shown in Figures 1 and 2 reveal important differences between 3-D and simpler models, which undermines confidence in previously published end-of-biosphere studies. It is beyond the scope of this study to attempt to determine the “best” model (or ensemble of models) for studying the future evolution of Earth's climate, although this could perhaps be an area of focus for a climate model intercomparison project (e.g., Sohl et al., 2024). For now, we will suffice to note that our 3-D calculations in this study represent a first attempt to map out the parameter space for Earth's future climate evolution with a complex model, which highlights the need for continued study of this problem with a hierarchy of models.

## 4. Habitability and Weathering Limits

Projections for the future evolution of Earth's climate and the end of the biosphere make two basic assumptions. First is that the incident solar flux will steadily increase as the sun brightens during the remainder of its main sequence phase. This assumption is fundamental to understanding of stellar evolution and is applied consistently across prior end-of-biosphere studies. Here, we follow the parameterized equation developed by Caldeira and Kasting (1992) for changes in solar luminosity with time:

$$\frac{S}{S_0} = \left(1 - 0.38\,\frac{t}{t_0}\right)^{-1}, \tag{1}$$

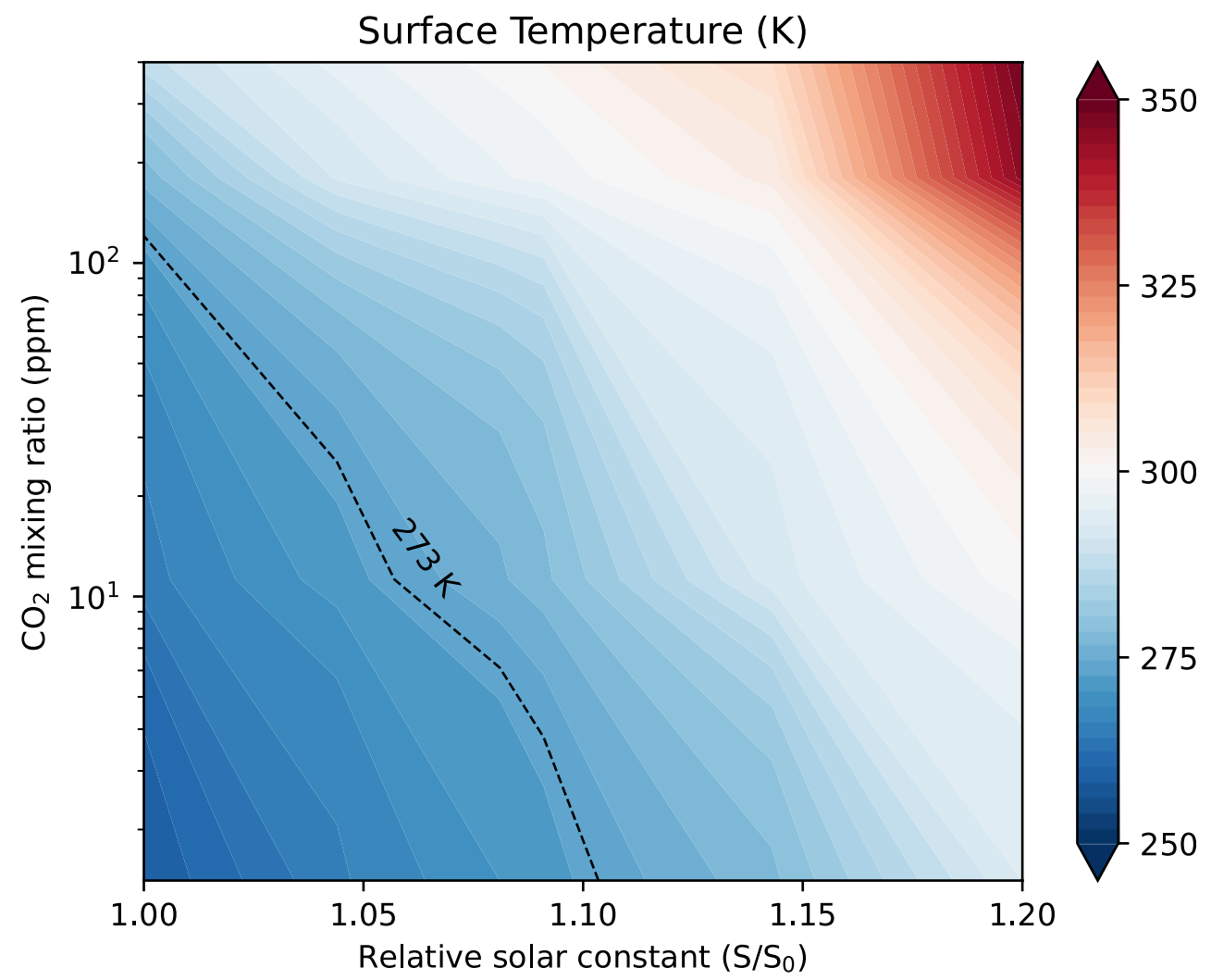

**Figure 3.** ExoCAM calculations of global average surface temperature as a function of relative solar constant and $CO_2$ mixing ratio. The dashed line indicates the freezing point of water.

where $t$ is time from the present in years, and $t_0 = 4.55$ Gyr. The second assumption is that the $CO_2$ mixing ratio will decrease in response to the enhancement of silicate weathering from a warmer climate as the sun brightens. Most previous studies have made the assumption that silicate weathering is strongly dependent on temperature, which should thereby drive decreases in $CO_2$ with time, although different choices for the functional representation and physical constants of the weathering models give different evolutionary trajectories for Earth's future climate. Furthermore, Graham et al. (2024) have demonstrated a weathering model in which the $CO_2$ mixing ratio does not monotonically decrease with time, which suggests the possibility that silicate weathering may have much weaker dependence on temperature than previously assumed. It is worth noting that the calculations by Graham et al. (2024) used a 0-D climate model, which may neglect factors such as precipitation, run-off, and continental location that may contribute to the silicate weathering rate (e.g., Baum et al., 2022; Jansen et al., 2019; Lehmer et al., 2020; Otto-Bliesner, 1995). In this study, we use a 3-D climate model to explore both possibilities at their extreme limits.

The full set of 29 ExoCAM simulations performed for this study are listed in Table 1. This simulation set shows changes in climate as solar constant progresses through time at present-day ($S/S_0 = 1.0$) and in the future at 0.5 Gyr ($S/S_0 = 1.044$), 0.9 Gyr ($S/S_0 = 1.081$), 1.0 Gyr ($S/S_0 = 1.091$), 1.5 Gyr ($S/S_0 = 1.044$), and 2.0 Gyr ($S/S_0 = 1.2$). Simulations are performed at each value of solar constant for a range of $CO_2$ mixing ratios from 400 ppm down to 1.4 ppm (and 0 ppm for the 2 Gyr case). These simulations otherwise assume an Earth-like atmosphere with 1 bar $N_2$ and 1.8 ppm $CH_4$, Earth-like topography, and a slab ocean (see Appendix C for details). The global average surface temperature for each simulation is listed in the fourth column of Table 1 and shown as a contour plot in Figure 3. The hot region in the upper-right corner of this surface temperature plot shows the end state for a planet that experiences increases in solar constant but no change in $CO_2$. Other regions of the parameter space indicate possible trajectories for the evolution of Earth's climate, depending on the extent to which temperature-dependent weathering processes can regulate global temperature.

Surface temperature by itself is an insufficient metric for describing changes in habitability due to a brightening sun and decrease in $CO_2$, and previous end-of-biosphere studies have all constructed functions to represent the aggregate productivity of the biosphere. In these models, productivity is linked to temperature and $CO_2$ mixing ratio, drawing on empirical findings on the behavior of Earth's vegetation, with additional limits on C3 and C4 photosynthesis when $CO_2$ mixing ratio falls below a critical threshold (typically this is 150 ppm for C3 photosynthesis and 10 ppm for C4 photosynthesis). Previous studies have indeed demonstrated successfully that the assumptions underlying the biological productivity model can alter the timing for any end-of-biosphere events. In this study, we set aside any $CO_2$-based limits on photosynthesis for now (returning to this idea in §5) and focus on understanding the changes in average and spatial habitability across our parameter space. Our analysis that follows includes use of a habitability metric that was developed by Woodward et al. (2025) for specific application to studies using 3-D climate models (described further in Appendix D).

We use the set of ExoCAM simulations listed in Table 1 and shown in Figure 3 to show snapshots of two possible trajectories for Earth's future climate evolution. We first show a trajectory that corresponds to the possibility that the temperature-dependence of silicate weathering is weak, as suggested by Graham et al. (2024), and we then show a second trajectory that corresponds to strong temperature-dependence of silicate weathering. This approach allows us to leverage our computationally expensive set of ExoCAM simulations to examine these two extreme possibilities for future climate evolution, without requiring an explicit numerical representation of the carbonate-silicate cycle. Although some studies have experimented with methods for using a 3-D climate model to simulate future climate evolution into a runaway greenhouse (e.g., Chaverot et al., 2023), such studies depend on simplifying the time-stepping process across these long timescales. For the present study, we refrain from attempting to bind ourselves toward any particular functional representation of silicate weathering or any choice of weathering rate values. We instead focus on examining the two possibilities of weak weathering and strong weathering as end member trajectories for Earth's climate evolution.

Future Earth, fixed $CO_2$
Present
$\Delta S/S_0$=0%
Mean: 288 K
$CO_2$: 400 ppm
0.5 Gyr Future
$\Delta S/S_0$=4.4%
Mean: 296 K
$CO_2$: 400 ppm
1.0 Gyr Future
$\Delta S/S_0$=9.1%
Mean: 301 K
$CO_2$: 400 ppm
1.5 Gyr Future
$\Delta S/S_0$=14%
Mean: 309 K
$CO_2$: 400 ppm
2.0 Gyr Future
$\Delta S/S_0$=20%
Mean: 348 K
$CO_2$: 400 ppm

**Figure 4.** ExoCAM simulations of Earth from the present up to 2.0 Gyr in the future, assuming $CO_2$ remains fixed at 400 ppm. The top row shows surface temperature, the middle row shows precipitation minus evaporation, and the bottom row shows the habitability metric developed by Woodward et al. (2025) (Equations D1 and D2). All quantities are 20-year averages after the model has reached a steady-state. Increases in temperature due to a brightening sun cause an increase in habitability up to 1.5 Gyr due to the greening of the poles, but habitability decreases significantly by 2.0 Gyr as Earth becomes too hot for complex vegetative life. This corresponds to a weak temperature dependence of weathering, in which Earth's $CO_2$ levels do not change even as the climate warms (cf., Graham et al., 2024).

The trajectory analogous to weak weathering is shown in Figure 4, which shows the time evolution of Earth's climate from present to 2.0 Gyr in the future with the assumption that $CO_2$ mixing ratio remains fixed at 400 ppm. As the sun brightens by a total of 20% over this 2.0 Gyr duration, Earth's surface temperature (top row) begins to increase—slowly at first up to about 1.5 Gyr, with a gradual depletion of the ice caps, and then much greater warming by 2.0 Gyr with a significantly reduced surface temperature gradient. The balance of precipitation and evaporation (middle row) shows an increase in maximum precipitation out to about 1.5 Gyr, with the greatest values near the equator, followed by a decrease with the regions of maximum precipitation shifted toward the poles. Surface habitability (bottom row) increases in the future simulations up to 1.5 Gyr, with most of the increases in complex habitability at polar latitudes as the ice caps deplete; however, by 2.0 Gyr, the regions of complex habitability have vanished and only microbial-dominated environments remain (see Appendix D for details on the habitability metric). These time slices of indicate a theoretical limit in which silicate weathering is weakly dependent on temperature, to the extent that the carbonate-silicate cycle does not significantly change the distant future value of $CO_2$ mixing ratio compared to today (cf., Graham et al., 2024). In this case, we have fixed the $CO_2$ mixing ratio at present-day values as a constant with time; this is most certainly an oversimplification of how $CO_2$ should evolve, even if weathering is weak, but it represents an end-member trajectory for the evolution of Earth's biosphere when net $CO_2$ exchange is zero.

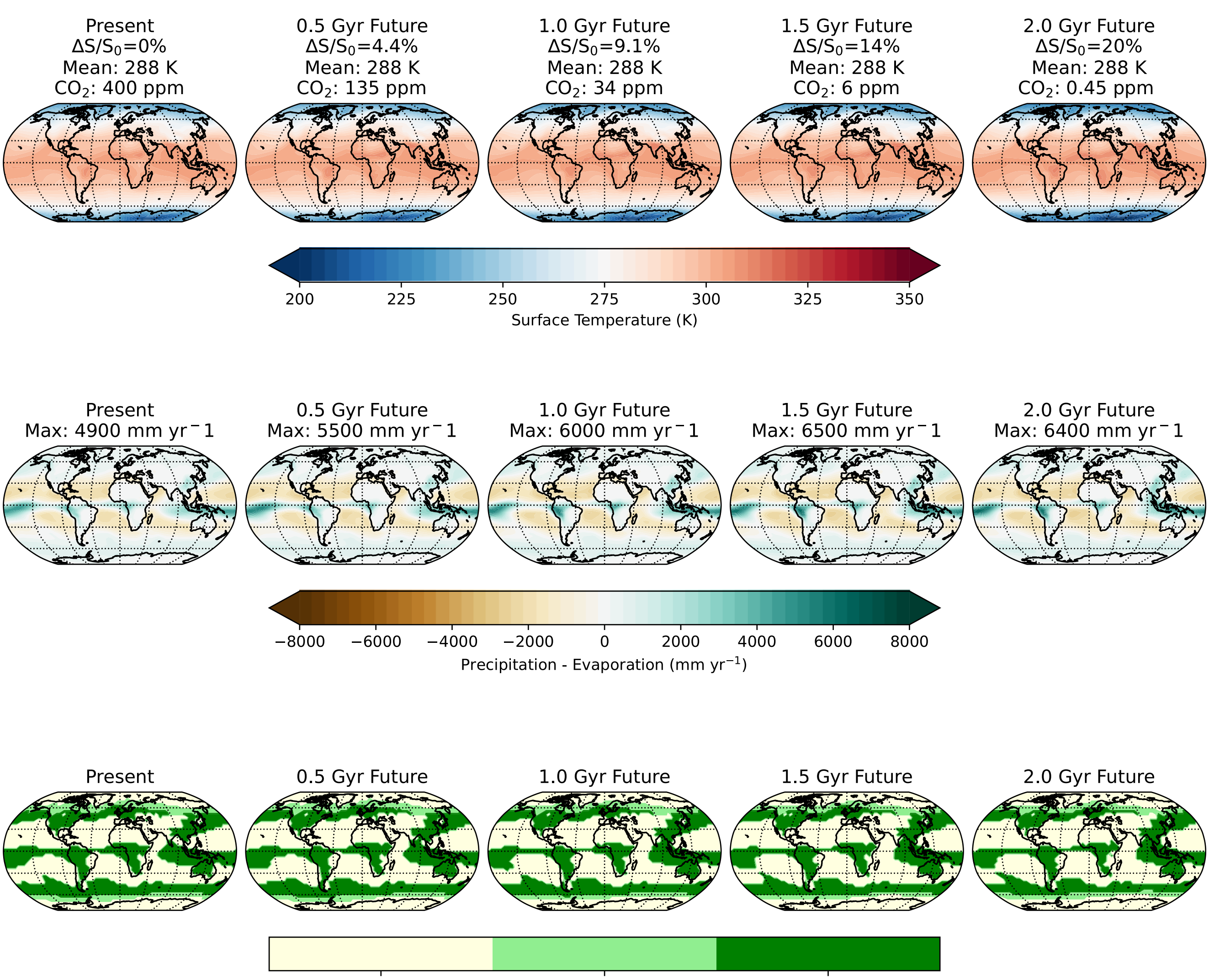

**Figure 5.** ExoCAM simulations of Earth from the present up to 2.0 Gyr in the future, assuming global average temperature remains fixed at 288 K. The top row shows surface temperature, the middle row shows precipitation minus evaporation, and the bottom row shows the habitability metric developed by Woodward et al. (2025) (Equations D1 and D2). All quantities are 20-year averages after the model has reached a steady-state. This corresponds to a strong temperature dependence of weathering, in which the carbonate-silicate cycle cause Earth's $CO_2$ levels decrease as insolation increases in order to maintain a constant global average temperature. The habitability metric does not change significantly over this 2.0 Gyr span; however, C3 photosynthesis would cease at a $CO_2$ mixing ratio of ∼150 ppm and C4 photosynthesis would end at ∼10 ppm (cf., Caldeira & Kasting, 1992). A biosphere dominated by CAM photosynthesis might be possible at lower $CO_2$ levels.

The trajectory analogous to strong weathering is shown in Figure 5, which shows the time evolution of Earth's climate from present to 2.0 Gyr in the future with the assumption that global average surface temperature remains a constant 288 K. As the sun brightens by a total of 20% over this 2.0 Gyr duration, Earth's $CO_2$ mixing ratio decreases at a rate that preserves a constant present-day value of global average surface temperature. The spatial distribution of temperature (top row) shows only small variations at the different snapshots, with the ice caps remaining present up through 2.0 Gyr. The balance of precipitation and evaporation (middle row) shows an increase in maximum precipitation that peaks at about 1.5 Gyr, but the overall precipitation pattern does not significantly change. Surface habitability (bottom row) shows only minor differences in the distribution pattern, with the poles remaining in a state of limited habitability even with a 20% brighter sun at 2.0 Gyr. However, these habitability metrics at 1.5 Gyr and later may be misleading if C4 photosynthesis would be $CO_2$-starved below 10 ppm (cf., Caldeira & Kasting, 1992); we examine this question of $CO_2$ limits for vegetative life in §5. These time slices indicate a theoretical limit in which silicate weathering is strongly dependent on temperature, to the extent that the carbonate-silicate cycle will draw down sufficient $CO_2$ from the atmosphere to maintain a constant value of global average surface temperature. In this case, we have represented such a trajectory by conducting simulations at fixed values of $CO_2$ that correspond to a 288 K global average temperature; this again is an oversimplification of $CO_2$ (and thus temperature) evolution under strong weathering, but it represents an end-member trajectory for the evolution of Earth's biosphere when net $CO_2$ exchange maintains a constant global average temperature.

The simulations shown in Figures 4 and 5 are illustrative of weak and strong weathering trajectories that can place limits on the lifetime of the vegetative biosphere (discussed next in §5). It is worth emphasizing that these are all steady-state simulations, rather than an attempt at using a 3-D model to simulate a 2.0 Gyr time evolution. Likewise, we do not include any explicit feedback in our models between the climate and the carbonate-silicate cycle. These results should therefore be interpreted as illustrative of end-member trajectories for Earth's future climate, which could serve as motivation for subsequent studies that attempt a more rigorous coupling between 3-D climate and carbonate-silicate models. Nevertheless, these results demonstrate that both weak and strong weathering trajectories show the possibility for Earth's complex habitability to persist long into the distant future.

## 5. Lifetime of the Biosphere

We now return to the primary question motivating this study: what is the maximum lifetime of the vegetative biosphere? We focus on the "vegetative" biosphere because $CO_2$ limits for productivity are specific to photosynthesis. The thermal limits for life apply to plants as well as other animals, but some possibilities (including technology) could allow forms of life to persist even if the surface of Earth rises above these temperatures (discussed further in §6). We also focus on estimating the "maximum" lifetime of the vegetative biosphere, acknowledging that our steady-state ExoCAM simulations showing the weak and strong weathering cases (Figures 4 and 5) are representations of end-member trajectories; actual trajectories for Earth's climate may fall between these two extremes. With this in mind, we use on our calculations from Figures 4 and 5 to draw schematic trajectories with time for both the strong and weak weathering limits in Figure 6.

The weak weathering trajectory shown in Figure 6 (red line) illustrates the extreme case in which Earth's climate warms (top) while $CO_2$ remains fixed at present-day values (middle) in response to the brightening sun (bottom). This trajectory reaches the 323 K thermal limit at 1.68 Gyr, which occurs later than the 1.5 Gyr predicted by Caldeira and Kasting (1992) (gray diamond). The higher thermal limit of 338 K occurs at 1.87 Gyr, which agrees with the calculations by Graham et al. (2024). If these two thermal limits are taken as the possible maxima for plant life on Earth, then the maximum lifetime of Earth's vegetative biosphere under weak weathering is 1.68–1.87 Gyr.

The strong weathering trajectory shown in Figure 6 (violet line) illustrates the extreme case in which Earth's climate maintains present-day surface temperatures (top) while $CO_2$ decreases due to enhanced silicate weathering (middle) in response to the brightening sun (bottom). This trajectory reaches the C4 photosynthesis limit at 1.35 Gyr, which is later than the 0.9 Gyr predicted by Caldeira and Kasting (1992) (gray diamond) as well as predictions by other studies (gray X's). The canonical story typically stops here, with the assumption that a biosphere at a $CO_2$ mixing ratio below 10 ppm cannot sustain a photosynthetic biosphere (e.g., Caldeira & Kasting, 1992) and will transition at this point into a planet dominated by microbial life in polar and high altitude refugia (e.g., O'Malley-James et al., 2013; O'Malley-James et al., 2013). If these are firm limits, then the maximum lifetime of Earth's vegetative biosphere under weak weathering is 1.35 Gyr. However, we suggest that the 10 ppm $CO_2$ mixing ratio limit for photosynthesis is not a hard limit, and we discuss some possibilities that would enable vegetative life to persist at even lower $CO_2$ levels.

First, the study by Graham et al. (2024) favored a 2.9 ppm $CO_2$ mixing ratio threshold for C4 photosynthesis, and noted that this may even be a conservative limit. All previous studies have relied on a $CO_2$ mixing ratio of 10 ppm as a threshold for C4 photosynthesis, but observations of plants on Earth provide numerous examples of species that can thrive below this limit (e.g., Chen et al., 1970). If we adopt 2.9 ppm as the $CO_2$ threshold for C4 photosynthesis, then this extends the maximum lifetime of the vegetative biosphere to 1.64 Gyr.

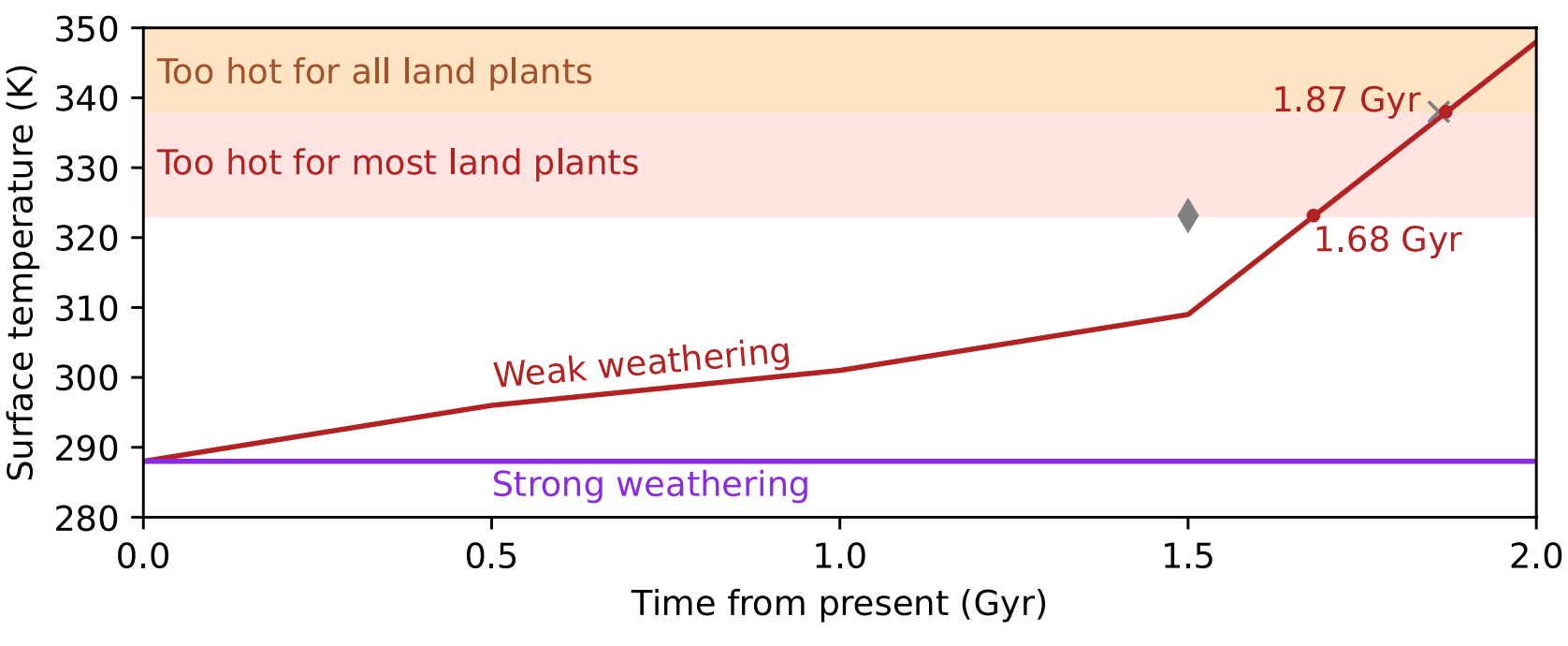

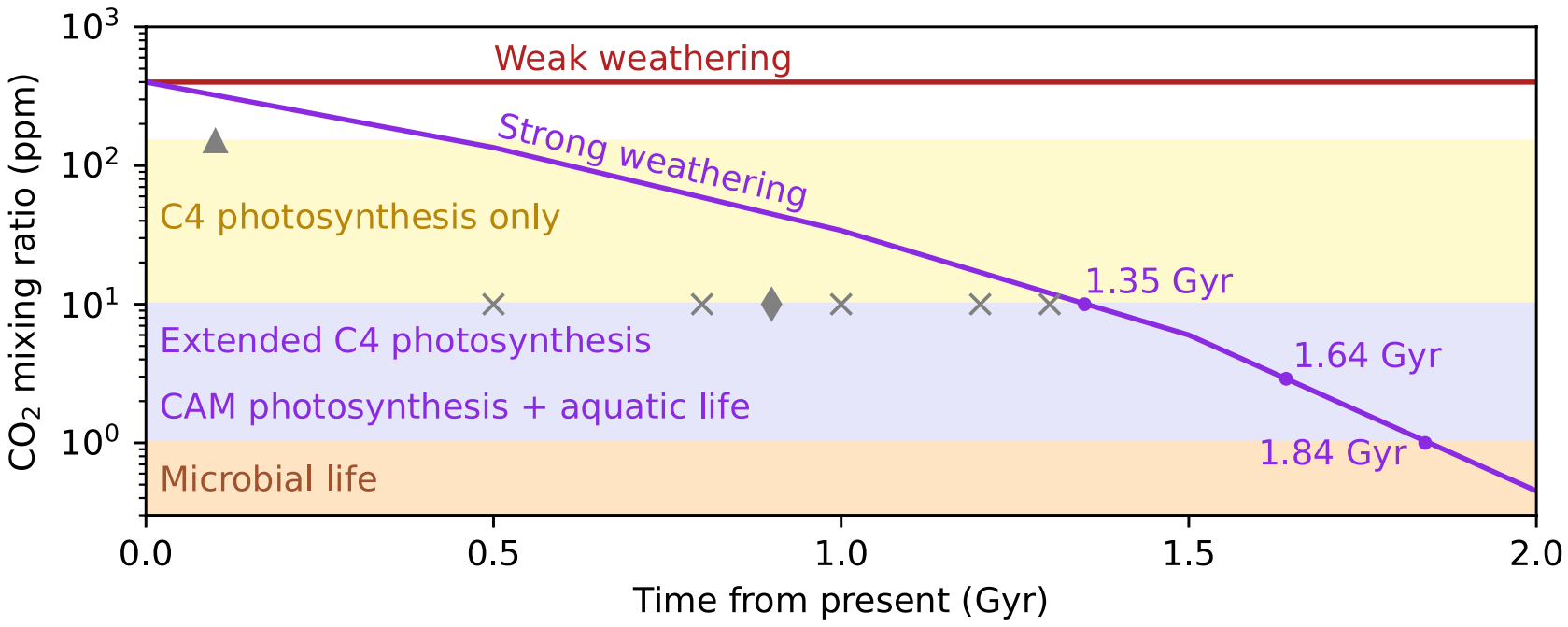

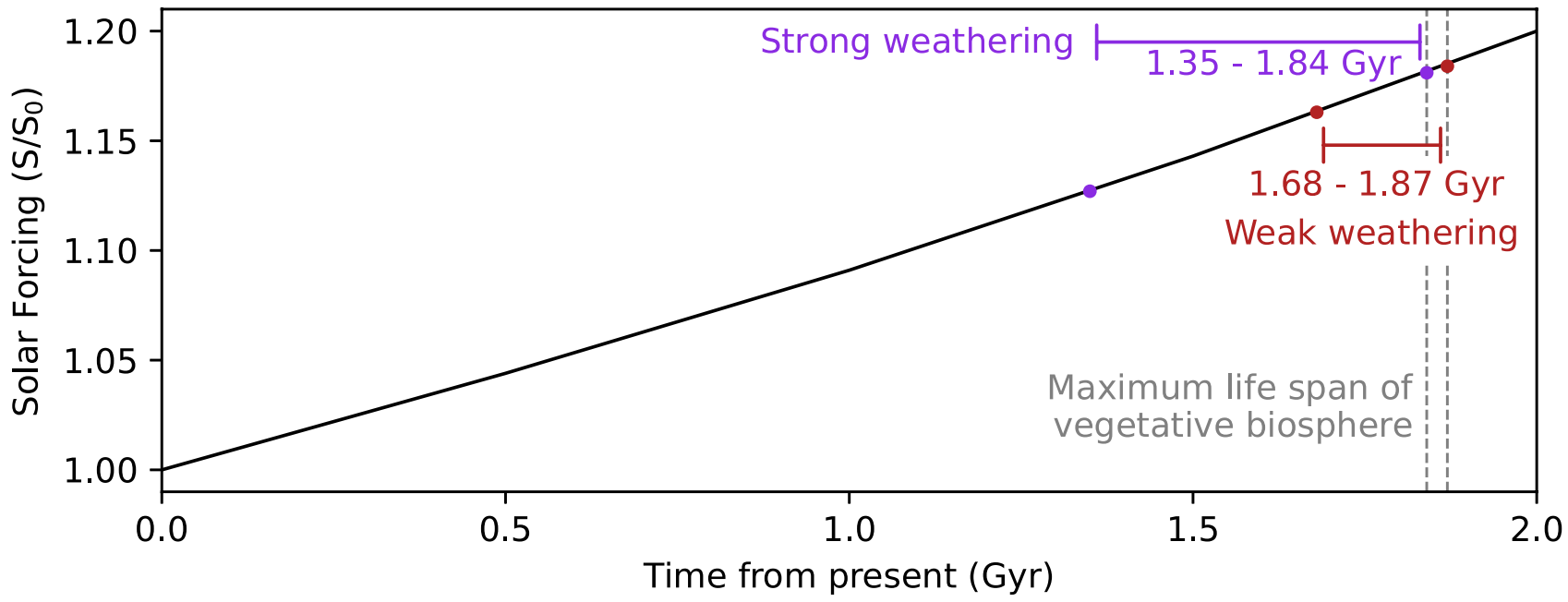

**Figure 6.** Schematic trajectories for the climate of Earth from the present up to 2.0 Gyr in the future, based on the ExoCAM simulations shown in Figures 4 and 5. The top panel shows surface temperature as a function of time. The weak weathering limit (red) shows increasing temperatures that eventually become too hot for most land plants (the traditional limit of 323 K (cf., Caldeira & Kasting, 1992)) at 1.68 Gyr and eventually too hot for all land plants (the optimistic limit of 338 K (cf., Graham et al., 2024)) at 1.87 Gyr. The strong weathering limit (violet) maintains a constant surface temperature. The middle panel shows $CO_2$ mixing ratio as a function of time. The strong weathering limit shows decreasing $CO_2$ levels that eventually inhibit both C3 and C4 photosynthesis at 1.35 Gyr at mixing ratios below 10 ppm. A biosphere dominated by CAM photosynthesis might be possible at $CO_2$ mixing ratios as low as 1 ppm, which could extend the habitable duration of a CAM biosphere to 1.84 Gyr. The weak weathering limit maintains a constant $CO_2$ mixing ratio. The bottom panel shows solar forcing as a function of time, which follows the expression in Equation 1. Gray markers indicate results from previous studies of the lifetime of the biosphere, with a triangle showing the estimate by Lovelock and Whitfield (1982), diamonds showing the estimates by Caldeira and Kasting (1992), and X marking estimates from other studies that were tabulated by Graham et al. (2024).

Second is the possibility that crassulacean acid metabolism (CAM) plants and other communities of plants may be able to sustain low carbon dioxide concentrations due to natural selection of already placed gas exchange stress strategies. Examples of CAM plants include cacti, agave, and some orchids. Some of these already in place strategies can be observed in obligate CAM plants (plants that are unable to shift between the CAM fixation strategy to C3 or C4) that are classified as strong CAMs or weak CAMs. Strong CAM plants are much more efficient when compared to weak CAMs in capturing $CO_2$, even in small quantities, and transform them into organic acids to later be broken down and utilized in the light-independent cycle of photosynthesis. An even more promising strategy in CAM plants, known as CAM-cycling, is where such plants recycle the $CO_2$ that is produced by the respiration process, thus having little need for atmospheric $CO_2$ (although some is still necessary to complete their photosynthetic process). The possibility that CAM photosynthesis could extend the lifetime of Earth's biosphere was noted by Graham et al. (2024); however, the authors did not include CAM plants in their analysis due to the lack of data for the responses of CAM plants to changes in $CO_2$ and temperature. Such laboratory work would be valuable for providing constraints to models of large-scale biological productivity for present-day and future Earth scenarios. For now, we note the possibility that CAM photosynthesis could continue past the threshold for C4 photosynthesis. We tentatively assign the CAM limit for $CO_2$ starvation at 1 ppm, which may even be a conservative limit.

Third, we note that diatoms and aquatic macrophytes do not only rely on atmospheric $CO_2$ and are able to use dissolved bicarbonate in their aqueous environment as a primary or secondary source (e.g., Burkhardt et al., 2001; Nimer et al., 1997; Tortell et al., 1997). If weathering in the carbonate-silicate cycle is temperature-dependent, then enhanced weathering due to a brightening sun would reduce $CO_2$ in the atmosphere while increasing the concentration of bicarbonate ions in seawater. This suggests the possibility that a planet with a $CO_2$-starved atmosphere under a strong weathering scenario may still support thriving communities of macroscopic organisms in bicarbonate-rich aquatic environments. Defining such a limit in terms of atmospheric $CO_2$ is difficult without further laboratory studies that explore the dependence of bicarbonate uptake by organisms at increasing temperatures and decreasing $CO_2$. We tentatively use 1 ppm $CO_2$ to denote this limit, recognizing that aquatic life may be able to thrive even below this threshold if weathering is sufficiently strong to saturate the ocean with bicarbonate ions.

Finally, it is important to recognize that photosynthetic ecosystems will inevitably be subject to evolutionary processes over the next 1–2 Gyr, and adaptions could occur that enable some vegetative communities to survive at lower $CO_2$ and/or higher temperatures than today. Some research has explored opportunities for utilizing conventional breeding, synthetic biology, and other methods for improving photosynthetic efficiency (e.g., Evans, 2013; Zhu et al., 2008; Zhu et al., 2010); such possibilities demonstrate the idea that plant evolution could conceivably remain open to developments that improve adaptation of the biosphere to a brightening sun. We do not attempt to speculate on the direction of any evolutionary trends for the vegetative biosphere but only note that we cannot necessarily dismiss the ability for life to adapt to changes and co-evolve with the planet.

In summary, our strong weathering limit ranges from a conservative $CO_2$ starvation threshold of 10 ppm to a lower value of 1 ppm. If these two productivity limits are taken as the possible maxima for plant life on Earth, then the maximum lifetime of Earth's vegetative biosphere under strong weathering is 1.35–1.84 Gyr. We note that the highest value of this range for the strong weathering lifetime corresponds approximately to the highest value in the range for the weak weathering lifetime, which average to 1.86 Gyr. This is substantially longer than all previous studies (except Graham et al. (2024)) have calculated, and it suggests the possibility that Earth's photosynthetic biosphere could remain viable in some form up to the point at which Earth begins to lose its water. If this is the case, then the maximum lifetime of Earth's vegetative biosphere is comparable to the lifetime of Earth's oceans.

## 6. The Fate of Earth

The results of this study suggest that vegetative life on Earth can persist up until the planet loses its water; however, models give a wide range of predictions for when this should occur. The “moist greenhouse” limit describes the point at which the oceans are lost due to photodissociation of water followed by the escape of hydrogen to space. The 0-D model by Caldeira and Kasting (1992) estimated the onset of the moist greenhouse at ~1.5 Gyr when surface temperatures reach ~350 K and the loss of all water from the surface at ~2.5 Gyr, which would cause the carbonate-silicate cycle to cease, thereby allowing the accumulation of $CO_2$ to drive Earth into a hot Venus-like state. The 3-D ExoCAM simulations by Wolf and Toon (2014) found stable Earth climates at solar constant values as high as $S/S_0 = 1.155$, with mean surface temperatures at or below ~312 K, which suggested that the moist greenhouse limit may not occur for at least 1.5 Gyr or longer. A subsequent study by Wolf and Toon (2015) using the same 3-D model found an abrupt climate shift to a moist greenhouse regime at $S/S_0 = 1.125$ but observed that diffusion-limited water loss became significant at $S/S_0 = 1.19$ when mean surface temperatures reach ∼350 K; the authors concluded that this process would render Earth uninhabitable for water-based life by ∼2.1 Gyr from now. The 1-D model calculations by Kasting et al. (2015) also found a transition to a moist greenhouse climate at a ∼350 K surface temperature. Only two of the ExoCAM simulations conducted for this present study (Table 1) reside near this ∼350 K moist greenhouse threshold: these are simulations at $S/S_0 = 1.2$ with 400 ppm $CO_2$ (top-of-atmosphere specific humidity of $1.2 \times 10^{-3}$ kg $kg^{-1}$) and 180 ppm $CO_2$ (top-of-atmosphere specific humidity of $2.2 \times 10^{-4}$ kg $kg^{-1}$). The first of these could lose the entirety of Earth's oceans to space in ∼3 Gyr (cf., Wolf & Toon, 2015, Section 4.1), while the second would likely avoid such a fate. Average surface temperature in the other simulations remains below 310 K. This suggests that the weak weathering limit will reach the ∼350 K moist greenhouse threshold at ∼2.1 Gyr (cf., Wolf & Toon, 2015), but the strong weathering limit may remain resilient against a moist greenhouse for much longer.

If Earth does not enter a moist greenhouse state, then the oceans could be lost through a "runaway greenhouse" process, in which a positive feedback between temperature and evaporation drives increases in the water vapor greenhouse that rapidly deplete the ocean. The ExoCAM simulations by Wolf and Toon (2015) found that Earth remains stable against a runaway greenhouse up to $S/S_0 = 1.21$ (∼2.19 Gyr), noting that the moist greenhouse would be more likely to occur first. These estimates would imply that the strong and weak weathering trajectories could reach their terminal states of $CO_2$ starvation and thermal limits, respectively, before Earth begins to lose its water. However, a study using a different 3-D model (LMD generic; Leconte et al., 2013) found the onset of the runaway to occur at a lower threshold of $S/S_0 \approx 1.1$ (∼1.1 Gyr); this would imply that the runaway greenhouse would be the first process to end the vegetative biosphere. This range of predictions for the timing of the runaway greenhouse reveals the limitations and gaps in knowledge regarding the behavior of terrestrial climate at high insolation and low $CO_2$. Nevertheless, even without strong quantitative constraints on the runaway greenhouse timing, the brightening sun will eventually trigger a runaway greenhouse that marks the end of the biosphere as we know it. Beyond this, some unicellular communities might be able to survive at high altitudes and polar latitudes up to 2.8 Gyr (O'Malley-James et al., 2013, 2014) until the brightening and evolving sun eventually sterilizes—and then engulfs—our planet.

This paper has focused on the lifetime of the *vegetative* biosphere in particular, given that animal life and nearly all other forms of complex life on Earth depend on vegetation as a primary or secondary source of energy. But non-vegetative life can also engage in other behaviors that could alter or extend Earth's habitable lifetime: most notably, human technology today offers a glimpse of the possibilities that a future technological civilization could use to prolong life on Earth. One option to avoid the effects of a brightening sun would be to utilize technology to offset these increases, which could include geoengineering by deploying reflective stratospheric aerosol or sunshades (e.g., Goldblatt & Watson, 2012). A longer-term solution could involve modifying Earth's orbit to a location that could remain habitable when the sun expands into a red giant (e.g., Birch, 1993). Another option to counteract the sun's evolution could be to remove (or "star-lift") mass from the sun to maintain a constant insolation and prolong its main sequence lifetime (e.g., Scoggins & Kipping, 2023). Such examples illustrate the possibility that future technology could prolong the lifetime of the biosphere, perhaps even beyond the main sequence phase of the sun.

In the absence of technology, life may still find a way to survive. Wordsworth and Cockell (2024) noted that there are no theoretical obstacles for the evolution and adaptation of photosynthetic life to a space environment. Wordsworth et al. (2025) demonstrated proof-of-concept of this idea using common biomaterials to enable the growth of eukaryotic green alga under Mars-like atmospheric conditions. Such technological demonstrations exemplify the possibility space open to future evolutionary developments. As an example, we can imagine a scenario in which plants evolve the ability to regulate their temperature and pressure, perhaps in response to changing climates. As the sun brightens, plants may favor an aerial environment and adapt accordingly, spreading to high-altitude terrain and into the stratosphere and beyond. From Earth's upper atmosphere, life could continue to disperse to low-gravity objects like comets and the moon as well as into free-floating space. Such a scenario in which life diffuses through space after Earth is gone challenges the very concept of a bio*sphere*. This is only an example, but the possibilities suggested by Wordsworth and Cockell (2024) remind us that life does not necessarily need to remain bound to a planet, and future evolutionary developments could still extend life beyond Earth even if the planet itself is rendered uninhabitable.

## 7. Conclusion

In this study, we have used 3-D climate calculations to show that the maximum lifetime of Earth's vegetative biosphere is ~1.86 Gyr from now. If silicate weathering is weakly dependent on temperature, then this lifetime is based on a 338 K thermal limit for complex life. If this thermal limit is instead taken at 323 K, then the maximum lifetime under weak weathering is 1.68 Gyr. If silicate weathering is strongly dependent on temperature, then this lifetime is limited by $CO_2$ starvation. In this study, we use a $CO_2$ starvation threshold of 1 ppm to account for the possibility of CAM photosynthesis at low $CO_2$ mixing ratios and the ability of aquatic macrophytes to utilize dissolved bicarbonate instead of atmospheric $CO_2$. If this $CO_2$ starvation limit is instead taken at the traditional 10 ppm, then the maximum lifetime under strong weathering is 1.35 Gyr. All of these estimates are longer than those predicted by most 1-D models. The moist or runaway limits for Earth may occur before this ~1.86 Gyr maximum lifetime, but modeling studies show a wide range of predictions for when these should occur.

Life on Earth is resilient, and limits posed by thermal stress or $CO_2$ starvation may only reflect our observations of the biosphere today rather than hard limits on how the biosphere may evolve. We acknowledge that the results of this study should be examined with other 3-D models, and that a community effort that compares model results at high insolation and low $CO_2$ would be the best way to constrain these timescales. In the absence of a more robust model intercomparison study, we suggest that the default story for our planet's future is that life will survive at least as long as Earth.

## Appendix A: Zero-Dimensional (0-D) Energy Balance Model

The model developed by Caldeira and Kasting (1992) calculates surface temperature, $T$, based on the energy balance between incoming solar radiation, $S$, and outgoing infrared radiation, which neglects variations in altitude, latitude, and longitude. This energy balance equation is written as

$$(1-\alpha)\frac{S}{4} = \sigma T_{\text{eff}}^4, \tag{A1}$$

with

$$T = T_{\text{eff}} + \Delta T, \tag{A2}$$

where $\alpha$ is planetary albedo, $\sigma$ is the Stefan-Boltzmann constant, $T_{\text{eff}}$ is the effective blackbody radiation temperature of Earth, and $\Delta T$ is a greenhouse warming factor. Expressions for $\alpha$ and $\Delta T$ were developed by performing least-squares fits to a set of 143 radiative-convective climate calculations (with one dimension as a vertical column) using the model of Kasting et al. (1993), which varied surface temperature and the partial pressure of carbon dioxide $(p_{CO_2})$ in the range of 273 K $< T <$ 373 K and $10^{-8}$ bar $< p_{CO_2} < 10^{-2}$ bar. These expressions gave $\alpha$ as a function of $T$,

$$\alpha = 1.4891 - 0.0065979T + \left(8.567\times10^{-6}\right)T^2, \tag{A3}$$

and $\Delta T$ as a function of both $T$ and $\phi = \log_{10}(p_{CO_2})$,

$$\begin{aligned}\Delta T = &\ 815.17 + \left(4.895\times10^{7}\right)T^{-2} - \left(3.9787\times10^{5}\right)T^{-1}\\ &- 6.7084\psi^{-2} + 73.221\psi^{-1} - \left(3.0882\times10^{4}\right)T^{-1}\psi^{-1}.\end{aligned} \tag{A4}$$

These parameterizations neglect any changes in cloud cover with temperature and hold cloud forcing at a constant value of 14 W m$^{-2}$.

## Appendix B: One-Dimensional (1-D) Energy Balance Model

The model developed by Williams and Kasting (1997) and modified by Haqq-Misra and Hayworth (2022) calculates surface temperature, $T$, as a function of latitude, based on the energy balanced between incoming solar radiation, $S$, outgoing infrared radiation $F$, and diffusive energy transport between latitudinal bands. The model neglects variations in altitude and longitude. This energy balance equation is written as

$$C\frac{\partial T}{\partial t} = S(1-\alpha) - F + \frac{\partial}{\partial x}\left[D\left(1-x^2\right)\frac{\partial T}{\partial x}\right], \tag{B1}$$

where $\alpha$ is planetary albedo, $C$ is effective heat capacity per unit area, $x$ is the sine of latitude, and the parameter $D$ represents thermal conductivity. The rightmost term of Equation B1 represents the equator-to-pole energy transport across latitudinal bands in the model.

The two models of Williams and Kasting (1997) and Haqq-Misra and Hayworth (2022) make many similar or identical implementations of physical processes that are summarized here, with details available in the respective papers. The model is implemented with 18 latitudinal bands fixed 10 degrees apart. Geography is represented by assigning each latitudinal band a fractional value for ocean coverage, assuming the rest is land, with ice coverage calculated based on the temperature at each band. The value of $C$ is also weighted according to the respective coverage of land, ocean, and ice at each band. These surface properties are used to calculate a surface albedo at each band. The diffusive parameter $D$ is kept constant, as is the impact of clouds on the radiative energy budget. The function for $S$ depends on time, zenith angle, and obliquity in order to represent the seasonal cycle.

The implementation of $\alpha$ and $F$ are the primary differences between these two energy balance models. The model by Williams and Kasting (1997) used the climate model of Kasting et al. (1993) to develop high-order polynomial parameterizations for planetary albedo and outgoing infrared radiative flux. The expression for $\alpha$ fit the results of more than 24,000 radiative-convective model runs, which depends on $T$, $p_{CO_2}$ (the partial pressure of carbon dioxide), zenith angle, and surface albedo, over the range of 190 K $< T <$ 360 K and $10^{-5}$ bar $< p_{CO_2} <$ 10 bar (Williams & Kasting, 1997, Equations A9 and A10). The expression for $F$ fit the results of about 300 radiative-convective model runs, which depends on $T$ and $p_{CO_2}$, over the range of 190 K $< T <$ 380 K and $10^{-5}$ bar $< p_{CO_2} <$ 10 bar (Williams & Kasting, 1997, Equation A18).

The model by Haqq-Misra and Hayworth (2022) implemented a lookup table for interpolating the values of $\alpha$ and $F$, based on calculations performed with the radiative-convective model of Kopparapu et al. (2013) (an updated version of the Kasting et al. (1993) model). The lookup table for $\alpha$ includes results of nearly 35,000 radiative-convective model runs, which depends on $T$, $p$ (total surface pressure), $f_{CO_2}$ (the mixing ratio of carbon dioxide), zenith angle, and surface albedo; the lookup table for $F$ includes the results of about 1,750 radiative-convective model runs, which depends on $T$, $p$, and $f_{CO_2}$. These lookup table calculations were performed over the range of 190 K $< T <$ 370 K, 1 bar $< p <$ 11 bar, and $10^{-6} < f_{CO_2} < 0.9$ (Haqq-Misra & Hayworth, 2022; Appendix A). The lookup table is implemented as a hash table with a nearest neighbor search. Differences in $\alpha$ and $F$ between the polynomial fit by and lookup table methods are shown by Haqq-Misra and Hayworth (2022, Figures 1 and 2).

## Appendix C: Three-Dimensional (3-D) General Circulation Model

The ExoCAM model developed by Wolf et al. (2022), like other general circulation models, calculates temperature, winds, and other atmospheric variables on a gridded domain of latitude, longitude, and altitude, based on thermodynamic energy balance and momentum balance (through the Navier-Stokes equations), with additional physical parameterizations for the surface, clouds, convection, and other atmospheric processes. ExoCAM in particular is a branch of the Community Earth System Model (CESM), originally developed at the National Center for Atmospheric Research (NCAR), that has been extended by Wolf et al. (2022) for application to a broader range of atmospheric compositions than present-day Earth.

The primary modification to ExoCAM is its implementation of the separate ExoRT radiative transfer model, which is capable of handling a wide range of atmospheric compositions and pressures, including past and present Earth, dense carbon dioxide atmospheres, anodic atmospheres, and even hydrogen-dominated atmospheres. ExoRT is based on 2-stream radiative transfer calculations, using the correlated-$k$ method and spectroscopic data from the HITRAN database. The version of ExoRT used by Wolf and Toon (2015) and Wolf et al. (2018) was based on the HITRAN 2004 database, but ExoRT was upgraded in 2020 to use the HITRAN 2016 database. The differences between ExoCAM results arising from this upgrade is discussed by Wolf et al. (2022). The calculations in this study use the *n68equiv* version of ExoRT, which has a pressure range from $10^{-5}$ bar to 10 bar and a temperature range from 100 to 500 K.

The simulations in this study assume Earth values for planetary radius, gravitational acceleration, and orbital period. Eccentricity and obliquity are both set to zero. A solar spectrum with 68 bands is used with a solar constant of 1,360 W $m^{-2}$. The atmosphere is composed of 1 bar composed of variable amounts of $CO_2$ ranging from 400 ppm to 0, a fixed 1.8 ppm $CH_4$, and the remainder as $N_2$; water vapor pressure is added to this based on the Clausius-Clapeyron relation. Ocean albedo constants are set to 0.06, and land albedo constants are set to 0.30. The use of ExoCAM in this study involves a present-Earth configuration for topography (without any additional soil or vegetation parameterizations), with 40 layers in the vertical grid and a 4° × 5° horizontal resolution. ExoCAM uses a finite volume dynamical core, and the simulations in this study use a thermodynamic slab ocean. The CAM4 package is used for cloud and mass-flux convection calculations. Simulations were initialized from a cold start and run for 70 model years, with the average of the final 20 years used for analysis. The present-Earth value for the $CO_2$ mixing ratio was used as a "tuning" parameter, with 400 ppm as the value that gives a ∼288 K global average surface temperature for this model configuration.

## Appendix D: Habitability Metric

The habitability metric ($H$) that was developed by Woodward et al. (2025) is based on thermal limits for life on Earth as well as on surface water fluxes as a tracer of nutrient availability. The metric first defines a component, $H_T$, based on the dependence of surface temperature ($T_s$) alone:

$$H_T(T_s) = \begin{cases} \text{complex} & \text{if} \quad 273.15\,\text{K} \le T_s \le 323.15\,\text{K}, \\ \text{microbial} & \text{if} \quad 253.15\,\text{K} \le T_s \le 395.15\,\text{K}, \\ \text{limited} & \text{otherwise.} \end{cases} \tag{D1}$$

The thermal limits for complex life are based on the a conservative estimate for the habitats of poikilotherms, or cold-blooded animals, while the microbial thermal limits are based on observations of extremophilic life on Earth. The dependence of habitability on the availability of water is represented as the difference between precipitation ($P$) and evaporation ($E$), along with a minimum precipitation threshold:

$$H(T_s, P, E) = \begin{cases} H_T(T_s) & \text{if} \quad P - E \ge 0 \;\&\; P \ge 250\,\text{mm year}^{-1}, \\ \text{limited} & \text{otherwise.} \end{cases} \tag{D2}$$

Equation D2 describes a binary dependence of habitability on $P$ and $E$ as proxies for water availability and nutrient transport. Taken together with Equation D1, the value of $H$ describes an environment as being primarily dominated by either "complex" or "microbial" life, with the alternative being a "limited" state in which none of the criteria are satisfied. The authors of the habitability metric emphasized that this categorization is intended to be descriptive of the predominant forms of life in an environment, rather than a strict limitation on what can or cannot occur, noting that regions designated as "microbial" or "limited" may still host some complex life. The authors further emphasized that an environment designated as "complex" would also be inhabited by microbial life. This framework was validated by Woodward et al. (2025) against Earth observations and provides a robust predictor of habitability across the surface of Earth when compared with other methods. For present-day Earth, the Woodward et al. (2025) metric finds the global habitable fraction (land and oceans) is 59% for microbial life and 36% for complex life. We leverage this framework here to examine changes in habitability across our parameter space.

## Conflict of Interest

The authors declare no conflicts of interest relevant to this study.

## Availability Statement

Data and configuration files for all model simulations are archived on Zenodo: https://doi.org/10.5281/zenodo.16584870 (Haqq-Misra & Wolf, 2025).

**Acknowledgments**
Thanks to Rafael Loureiro and Robin Wordsworth for helpful conversations during the conception of this study. JHM gratefully acknowledges support from the NASA Exobiology program under award 80NSSC22K1632. ETW acknowledges support from the Virtual Planetary Laboratory funded via the NASA Astrobiology Institute Program Grants 80NSSC23K1398 and 80NSSC18K0829. Any opinions, findings, and conclusions or recommendations expressed in this material are those of the authors and do not necessarily reflect the views of their employers or NASA.